\documentclass[aps,prl,reprint,preprintnumbers,superscriptaddress,amsmath,amssymb,bibnotes,longbibliography]{revtex4-2}
\usepackage{graphicx}% Include figure files
\usepackage{dcolumn}% Align table columns on decimal point
\usepackage{amsmath,amssymb,bbold,bm}
\usepackage{epstopdf}
\usepackage{bm,multirow,threeparttable}% bold math
\usepackage{flushend}
\usepackage{lineno}

\newcommand{\bea}{\begin{eqnarray}}
\newcommand{\bean}{\begin{eqnarray*}}
\newcommand{\eea}{\end{eqnarray}}
\newcommand{\eean}{\end{eqnarray*}}
\newcommand{\bmx}{\begin{array}}
\newcommand{\emx}{\end{array}}

\usepackage[pdfstartview=FitH, CJKbookmarks=true, bookmarksnumbered=true, bookmarksopen=true, colorlinks, pdfborder=001, linkcolor=blue, anchorcolor=blue, citecolor=blue,urlcolor=blue]{hyperref}% add hypertext capabilities

\newcommand{\be}{\begin{equation}}
	\newcommand{\ee}{\end{equation}}

\newcommand{\serge}{CeRh$_{6}$Ge$_{4}$\ }
\newcommand{\cerge}{CeRh$_{6}$Ge$_{4}$\ }
\newcommand{\sergeCo}{Ce(Rh$_{1-x}$Co$_x$)$_6$Ge$_4$\ }
\newcommand{\TC}{$T_{\rm C}$\ }

\begin{document}

\renewcommand{\abstractname}{} 
\title{Critical fluctuations and conserved dynamics in a strange ferromagnetic metal}

\author{Jin Zhan}
\thanks{These authors contributed equally to this work.\\ Correspondence should be addressed to:}
\affiliation{Center for Correlated Matter and School of Physics, Zhejiang University, Hangzhou 310058, China}
\author{Yongjun Zhang}
\thanks{These authors contributed equally to this work.\\ Correspondence should be addressed to:} 
\affiliation{Hubei Key Laboratory of Photoelectric Materials and Devices, School of Materials Science and Engineering, Hubei Normal University, Huangshi 435002, China}
\author{Jiawen Zhang}
\affiliation{Center for Correlated Matter and School of Physics, Zhejiang University, Hangzhou 310058, China}
\author{Yu Liu}
\affiliation{Center for Correlated Matter and School of Physics, Zhejiang University, Hangzhou 310058, China}
\author{Zhiyong Nie}
\affiliation{Center for Correlated Matter and School of Physics, Zhejiang University, Hangzhou 310058, China}
\author{Yuxin Chen}
\affiliation{Center for Correlated Matter and School of Physics, Zhejiang University, Hangzhou 310058, China}
\author{Lin Jiao}
\affiliation{Center for Correlated Matter and School of Physics, Zhejiang University, Hangzhou 310058, China}
\author{Yashar Komijani}
\affiliation{Department of Physics, University of Cincinnati, Cincinnati, Ohio 45221, USA}
\author{Michael Smidman}
\email{msmidman@zju.edu.cn}
\affiliation{Center for Correlated Matter and School of Physics, Zhejiang University, Hangzhou 310058, China}
\author{Frank Steglich}
\affiliation{Center for Correlated Matter and School of Physics, Zhejiang University, Hangzhou 310058, China}
\affiliation{Max Planck Institute for Chemical Physics of Solids, 01187 Dresden, Germany}
\author{Piers Coleman}
\email{coleman@physics.rutgers.edu}
\affiliation{Department of Physics and Astronomy, Rutgers University, Piscataway, New Jersey 08854, USA}
\affiliation{Department of Physics, Royal Holloway, University
		of London, Egham, Surrey TW20 0EX, UK.}
\author{Huiqiu Yuan}
\email{hqyuan@zju.edu.cn}
\affiliation{Center for Correlated Matter and School of Physics, Zhejiang University, Hangzhou 310058, China}
\affiliation{Institute for Advanced Study in Physics, Zhejiang University, Hangzhou 310058, China.}
\affiliation{Institute of Fundamental and Transdisciplinary Research, Zhejiang University, Hangzhou 310058, China.}
\affiliation{State Key Laboratory of Silicon and Advanced Semiconductor Materials, Zhejiang University, Hangzhou 310058, China.}

\date{\today}

%Abstract
\begin{abstract}
	
		The origin of the strange metallic behavior observed in a wide range of quantum materials is an open challenge to condensed matter physics.
	Historically, strange metals were  uniquely associated with antiferromagnetic quantum critical points (QCPs), 
	but a new generation of materials reveals their  association with uniform order parameters, such as ferromagnetism, valley or nematic order, suggesting a deeper common denominator.  
		At a QCP, order parameter fluctuations are characterized by the dynamical critical exponent $z$, which quantifies the space-time scaling asymmetry. Here, we report the observation of a divergence in the Gr\"uneisen ratio at the QCP of  the strange-metal ferromagnet CeRh$_6$Ge$_4$ with  a dynamical critical exponent $z=3$, signaling that the underlying quantum singularity involves a conserved degree of freedom.  Yet the magnetization  of this easy-plane ferromagnet is not conserved. We argue that the  $z=3$ strange criticality requires a description beyond the Landau paradigm, proposing a link with the gauge modes of the small-to-large Fermi surface transition and the associated gauge charge of the delocalizing heavy electrons.
	
\end{abstract}
\maketitle

The strange metal state that develops near quantum phase
transitions is a long-standing mystery\,\cite{Phillips2022,Hartnoll2022}. These unusual metallic states exhibit fundamental departures from
the conventional Fermi liquid physics of metals, with a
{temperature dependence of the resistivity $\Delta\rho\sim T$ [$\Delta\rho=\rho(T)-\rho(0)$]} {consistent with a
	``Planckian'' scattering rate \cite{Legros2019,PhysRevLett.91.246405,YRSnature,Taupin2022} and a logarithmically divergent heat capacity coefficient $C/T\sim -\log T$ \cite{YRSnature,Ronning2005,Michon2019}.} Moreover, their close association with superconductivity suggests that the strange metal 
provides the highly entangled quantum fabric of unconventional
and high temperature superconductivity. 

Today, a new generation of
strange metals challenges our understanding of quantum materials. In early
experiments, strange metals were uniquely associated with
antiferromagnetic settings, as observed in heavy-fermion, high-temperature and iron-based superconductors. However, recent experiments  on heavy fermion materials \cite{Shen2020}, as well as iron selenide \cite{Licciardello2019,Jiang2023}, twisted bilayer graphene  and tungsten diselenide \cite{Cao2020,Zhao2023}, reveal that strange metals also develop in the vicinity of quantum phase transitions associated with fluctuating uniform order, such as spin and valley ferromagnetism or nematic order.  These developments motivate new inquiries into the role of the critical fluctuations in strange metal physics. 

The study of ferromagnetism as a quantum phase transition has a long legacy \cite{Hertz1976,Millis1993,Moriya2006}. Hertz considered the Stoner ferromagnetic (FM) instability of metals as a quantum critical point (QCP), and later works revealed that  a 3D Hertzian FM QCP would exhibit an anomalous $\Delta\rho \sim T^{5/3}$ resistivity characteristic of non-Fermi liquid behavior \cite{Stewart2001}. However, quantum criticality in a FM metal is predicted to be interrupted by a first order transition due to various non-analyticities originating from coupling between the soft modes of magnetization and the itinerant electron gas \cite{Belitz99,Brando2016,Belitz2002}.

The discovery of quantum criticality and strange metal behavior
in ferromagnetic Kondo lattices
\cite{Shen2020,Steppke2013} has launched a new debate. Since both the Planckian resistivity $\Delta\rho\sim T$ and continuous transition are ostensibly
inconsistent with a Hertzian transition, what then is the nature of the
critical point? Should we regard the quantum phase
transition as the development of critical spin polarization waves,
or does it involve a Kondo delocalization of the electrons?
Belitz and Kirkpatrick have 
argued that the inclusion of spin-orbit interactions in
non-centrosymmetric systems
will re-establish a continuous quantum phase
transition within the framework of Hertzian criticality \cite{Kirkpatrick2020}. 
This idea has spurred additional
theoretical debate \cite{Miserev2022} on whether higher order non-linearities
within the Hertzian picture might still give rise to a first order transition. 
An experimental method that can effectively delineate between a local
or an itinerant ferromagnetic quantum criticality \cite{Yamamoto2010b} is thus of
paramount importance for the unfolding understanding of strange
metals.

Generally, a QCP is distinguished by the development of
long-range correlations in both space and time.  At a critical point 
the correlation length $\xi$ diverges as a power-law of the tuning parameter $r$ according to
$\xi\sim r^{-\nu}$. The temporal
correlation time $\xi_{\tau }$ also diverges, but it does so at a
different rate, so that $\xi_{\tau }\sim (\xi)^{z}\sim r^{-\nu z}$,
where $z$ is the dynamical critical exponent. 
A classic experimental probe of the divergence in the temporal
correlation time is provided by the Gr\"uneisen
parameter $\Gamma_{c} = \beta /C_{P}$, which is the ratio of the volume thermal
expansion coefficient $\beta (T) = d\log V/dT$, where $V $ is the
volume, to
the specific heat $C_{P}$ at constant pressure.  
General
scaling arguments
\cite{Zhu2003,Kuechler2003,Gegenwart2016}, independent of microscopic
considerations, 
show that at a QCP the 
Gr\"uneisen parameter diverges with the same scaling dimension as the
inverse tuning parameter, i.e 
$\Gamma_{c}\sim r^{-1} \sim \xi_{\tau }^{1/\nu z}$. 
{Under the usual hyper-scaling assumption of quantum criticality}
\cite{Zhu2003,Kirkpatrick2015}, the correlation time is cut-off by
thermal fluctuations and scales with the Planck time, 
$\xi_{\tau }\sim \frac{\hbar }{T}$, from which it follows that $\Gamma_{c}\sim T^{-1/\nu z}$.  In this way, temperature-dependent
measurements of the Gr\"uneisen parameter can be used to characterize
a QCP.  Since theory provides firm predictions for the relationship between dynamical critical exponent and the lifetime of the underlying critical modes, a measurement of $\Gamma_c$ can therefore, shed light on the nature of the transition.
	
While the application of hydrostatic pressure to \serge allows for the
realization of a ferromagnetic QCP in a clean stoichiometric compound
without introducing disorder \cite{Shen2020}, only a limited range of experimental
probes can be utilized to probe quantum criticality under
pressure. Since measurements of the Gr\"uneisen parameter at the FM
QCP of \cerge can provide vital information about the nature of the
quantum criticality, this motivates the search for dopants that
tune the ferromagnetic order \cite{Zhang2022,Xu2021}, while preserving the intrinsic quantum critical behaviors. Although the substitution of Si for Ge is the most
obvious choice for inducing chemical pressure, upon suppressing
ferromagnetism with Si-doping there is no longer a $T$-linear
resistivity, and anomalous behaviors occur at the critical
concentration \cite{Zhang2022}. Here we instead dope the transition
metal site by synthesizing \sergeCo single crystals, where we find
that $5\%$ Co-doping can induce a FM QCP while still exhibiting the
same strange metal behaviors in the resistivity and specific heat.

\begin{figure}[t]
	\begin{center}
		\includegraphics[width=\columnwidth]{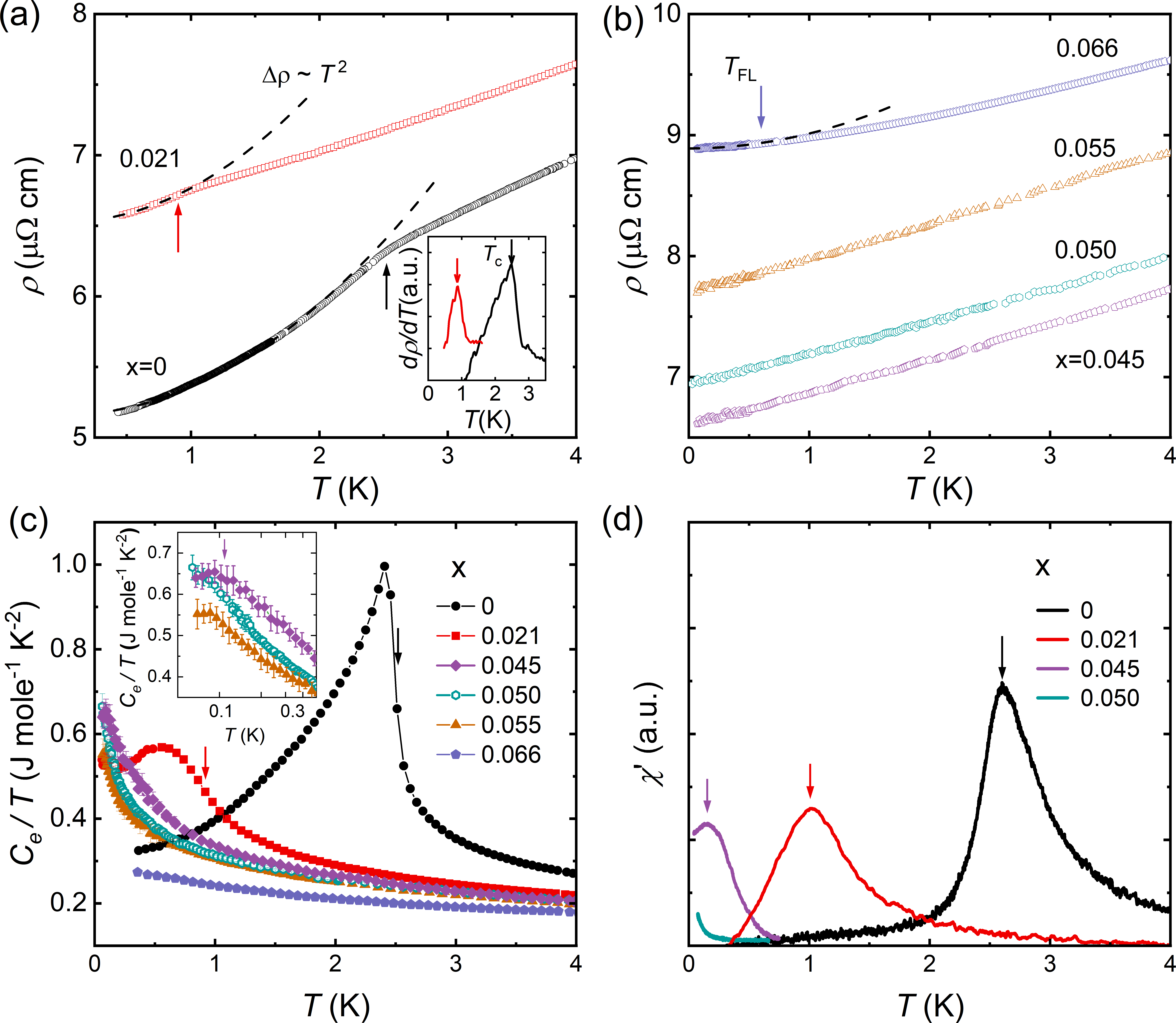}
	\end{center}
	\caption{Temperature dependence of the resistivity of Ce(Rh$_{1-x}$Co$_x$)$_6$Ge$_4$ with the current along the $c$ axis for (a)  $x\leq0.021$ and (b) $x\geq0.045$. The ferromagnetic transitions at \TC are detected in the $x=0$ and $x=0.021$ samples as shown by the vertical arrows, which for the resistivity are clearly observed in the derivative  in the inset of (a).  $\rho(T)$ for $x=0$ is shifted vertically by $3.6~\mu\Omega~{\rm cm}$ for clarity. The dashed lines show the $\Delta\rho\sim T^2$ behavior present below $T_{\rm FL}$. Temperature dependence of the (c) electronic specific heat coefficient $C_e/T$ (after subtracting the data of LaRh$_6$Ge$_4$), of {\sergeCo} single crystals down to 0.06~K, and (d) the real part of the ac susceptibility $\chi'$, where the ferromagnetic transition corresponds to a peak marked by vertical arrows.  The inset of (c) magnifies the low temperature $C_e/T$ below 0.5~K.}   \label{Fig1}\end{figure}

Single crystals of Ce(Rh$_{1-x}$Co$_x$)$_6$Ge$_4$ were grown using a Bi flux method \cite{Shen2020,Vosswinkel2012}. Figures~\ref{Fig1}(a) and \ref{Fig1}(b) display the low temperature electrical resistivity $\rho(T)$ of Ce(Rh$_{1-x}$Co$_x$)$_6$Ge$_4$, where the current is applied along the  $c$~axis. For $x=0.021$, \TC is detected at a lower
temperature of 0.9~K, most clearly in the derivative displayed in the inset. At higher concentrations ($x\geq0.045$), no transition is resolved in $\rho(T)$, and instead there is a linear temperature dependence, much like the strange metal behavior in pressurized \serge \cite{Shen2020}. On the other hand, transitions are
detected in the electronic specific heat as $C_e/T$ [Fig.~\ref{Fig1}(c)]
with $x=0.021$, as well as for $x=0.045$ when measuring down to 0.06~K. Meanwhile at a higher concentration of $x=0.050$ no transition is observed and $C_e/T$ diverges,  while for $x=0.055$, $C_e/T$ flattens at the lowest measured temperatures. Upon further increasing the concentration with $x=0.066$, there is the
emergence of $\Delta\rho\sim T^2$ behavior at low temperatures (dashed line in Fig.~\ref{Fig1}(b)), which together
with the flattening of $C_e/T$, signals the onset of a heavy Fermi
liquid state. Similarly, upon measuring the ac susceptibility
$\chi'$ down to 0.07~K  [Fig.~\ref{Fig1}(d)], the characteristic peak
corresponding to \TC is present at 0.15~K for the $x=0.045$ sample,
but is absent for $x\geq0.050$. Consequently, these results suggest
that Co-doping continuously suppresses \TC to a QCP at a critical
concentration $x_c\approx0.05$, at which there are the same strange
metal behaviors in $\rho(T)$ and $C_e/T$ as in the hydrostatic
pressure induced ferromagnetic QCP \cite{Shen2020}. Here the
difference from the Si-doped case may be due to the smaller disorder
effect, where for Co-doping $\rho(0)\approx 7~\mu\Omega~{\rm cm}$ at $x_c$,
as compared to $\rho(0)\approx 12~\mu\Omega~{\rm cm}$ for Si \cite{Zhang2022}.

\begin{figure}[t]
	\begin{center}
		\includegraphics[width=0.65\columnwidth]{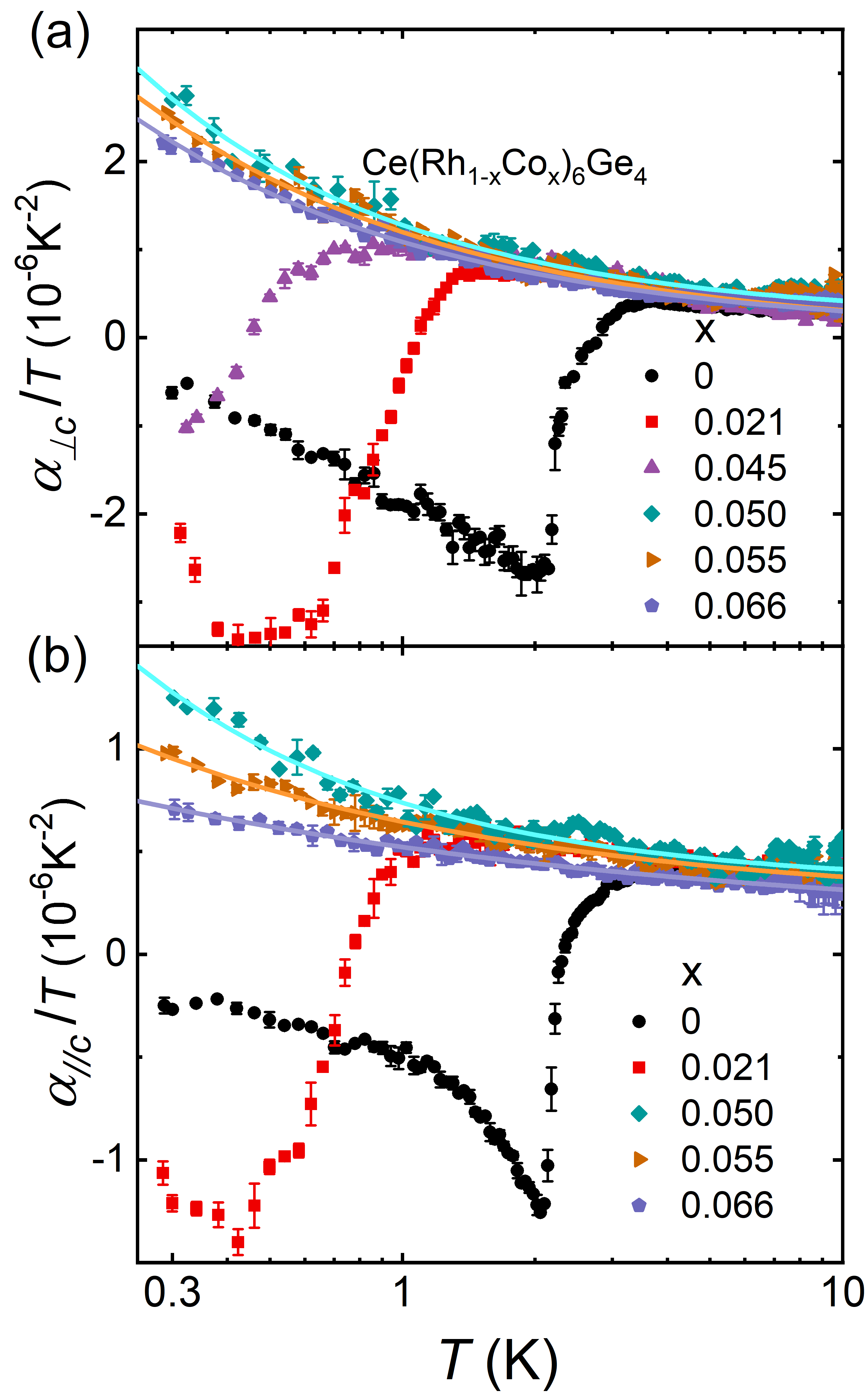}
	\end{center}
	\caption{(Color online) Temperature dependence of the linear thermal expansion coefficient (a) perpendicular ($\alpha_{\perp c}$), and (b) parallel ($\alpha_{\parallel c}$), to the $c$-axis.  For samples exhibiting magnetic transitions, there is a drop of $\alpha/T$ upon cooling for both directions. For $x=0.050$, 0.055, and 0.066,  $\alpha/T$  continues to increase with decreasing temperature, which can be described by a power law behavior  $\alpha/T\sim T^{-n}$  as shown by the solid lines, where the corresponding $n$ for $\alpha_{\perp c}$ are 0.70(6), 0.60(4), and 0.61(3), respectively, while for $\alpha_{\parallel c}$ they are 0.70(6), 0.47(6), and 0.33(5).}
	\label{Fig2}
\end{figure}

\begin{figure}[t]
	\begin{center}
		\includegraphics[width=0.7\columnwidth]{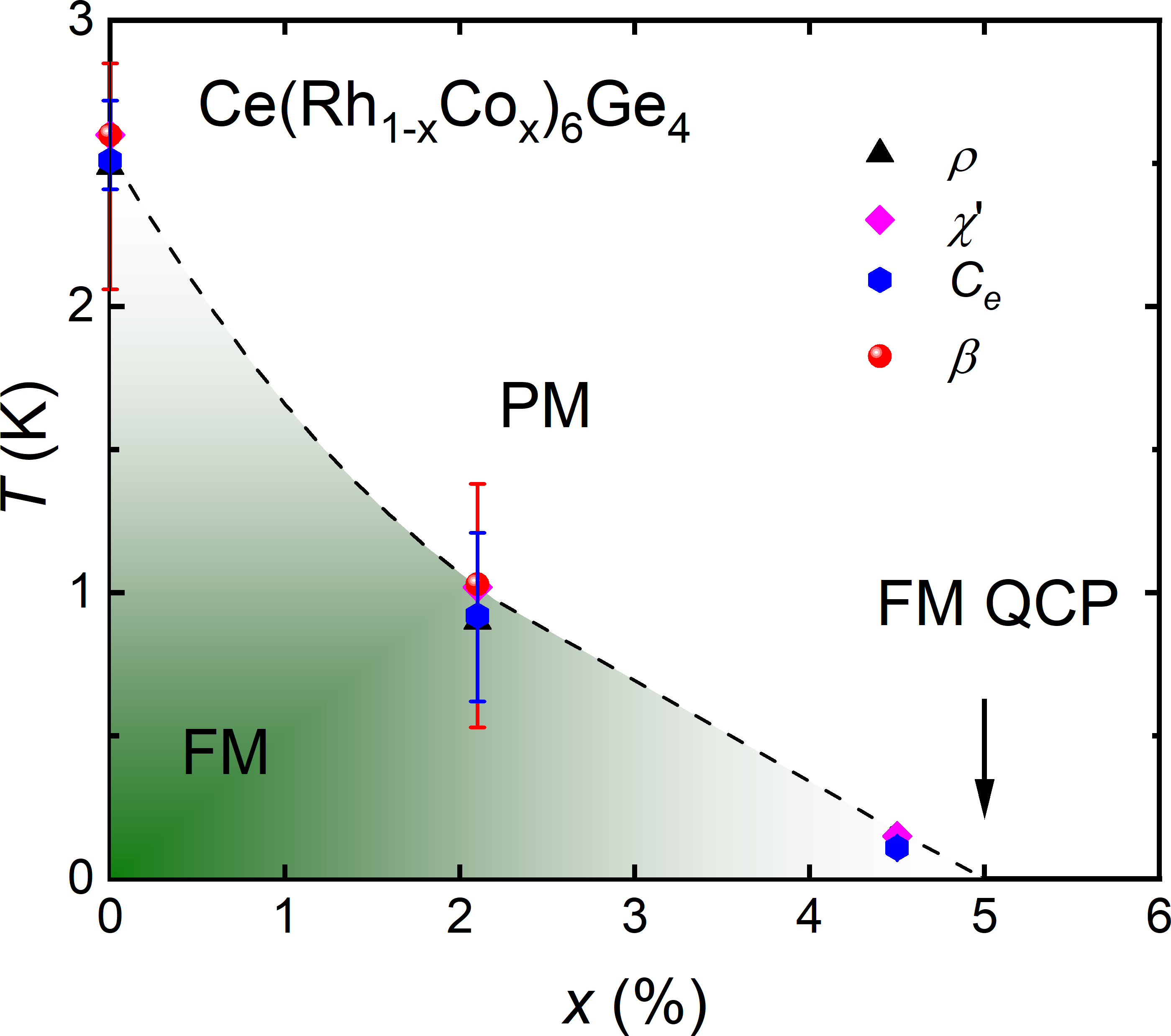}
	\end{center}
	\caption{Temperature-doping phase diagram of \sergeCo determined from  resistivity, ac susceptibility, specific heat, and thermal expansion measurements. The \TC for the specific heat and thermal expansion are determined from the midpoints and change of sign of $\beta(T)$, respectively, while the error bars correspond to the transition widths. The ferromagnetic transition is suppressed by doping, and is no longer detected for $x=0.050$, which is therefore in close proximity to a ferromagnetic QCP.}   \label{Fig3}\end{figure}

Given that Co-doping preserves the ferromagnetic quantum criticality and strange metal behavior, a wider range of probes are accessible at the QCP, such as the linear thermal expansion. The thermal expansion was measured parallel and perpendicular to the $c$ axis using high-resolution dilatometers in a $^3$He refrigerator down to 0.3~K \cite{Kuchler2012}.  The temperature dependence of the linear thermal expansion coefficient as $\alpha/T$, is displayed in Fig.~\ref{Fig2} both parallel ($\alpha_{\parallel c}$) and perpendicular ($\alpha_{\perp c}$) to the $c$ axis down to 0.3~K. Above around 3~K, there are similar positive values for all concentrations, but upon cooling there is a pronounced drop in those samples exhibiting magnetic transitions, which crosses to negative $\alpha/T$ for both directions, before reaching a sharp minimum. For  $x=0.045$, the onset of the drop in $\alpha/T$ is observed, which is a precursor to the transition  occuring at  lower temperatures [Figs.~\ref{Fig1}(c) and ~\ref{Fig1}(d)]. Meanwhile  markedly different behavior  is found for  samples with $x\geq x_c$, whereby $\alpha/T$ continues to increase  upon cooling. Above the critical concentration, $\alpha/T$ diverges with a power law dependence $\alpha/T\sim T^{-n}$, where for $x=x_c$, $n\approx0.7$ for both directions. Note also the magnitude of $\alpha_{\perp c}$ is consistently two-three times larger than  $\alpha_{\parallel c}$, which may also reflect the anisotropy of the crystalline-electric field ground state and corresponding quasi-1D hybridization \cite{Shu2021,Wu2021}. Figure~\ref{Fig3}  shows the temperature-$x$ phase diagram for \sergeCo obtained from combining these different measurements. Here the \TC from  thermal expansion results are determined from where there is a sign change of the volume thermal expansion coefficient $\beta=2\alpha_{\perp c}+\alpha_{\parallel c}$ \cite{signchange,SI}, and there is good agreement with the \TC obtained from $\rho(T)$, $C_e/T$, and $\chi'$. Together these strongly suggest that Co-doping suppresses \TC to a ferromagnetic QCP at $x_c\approx 0.05$.

\begin{figure}[t]
	\begin{center}
		\includegraphics[width=0.7\columnwidth]{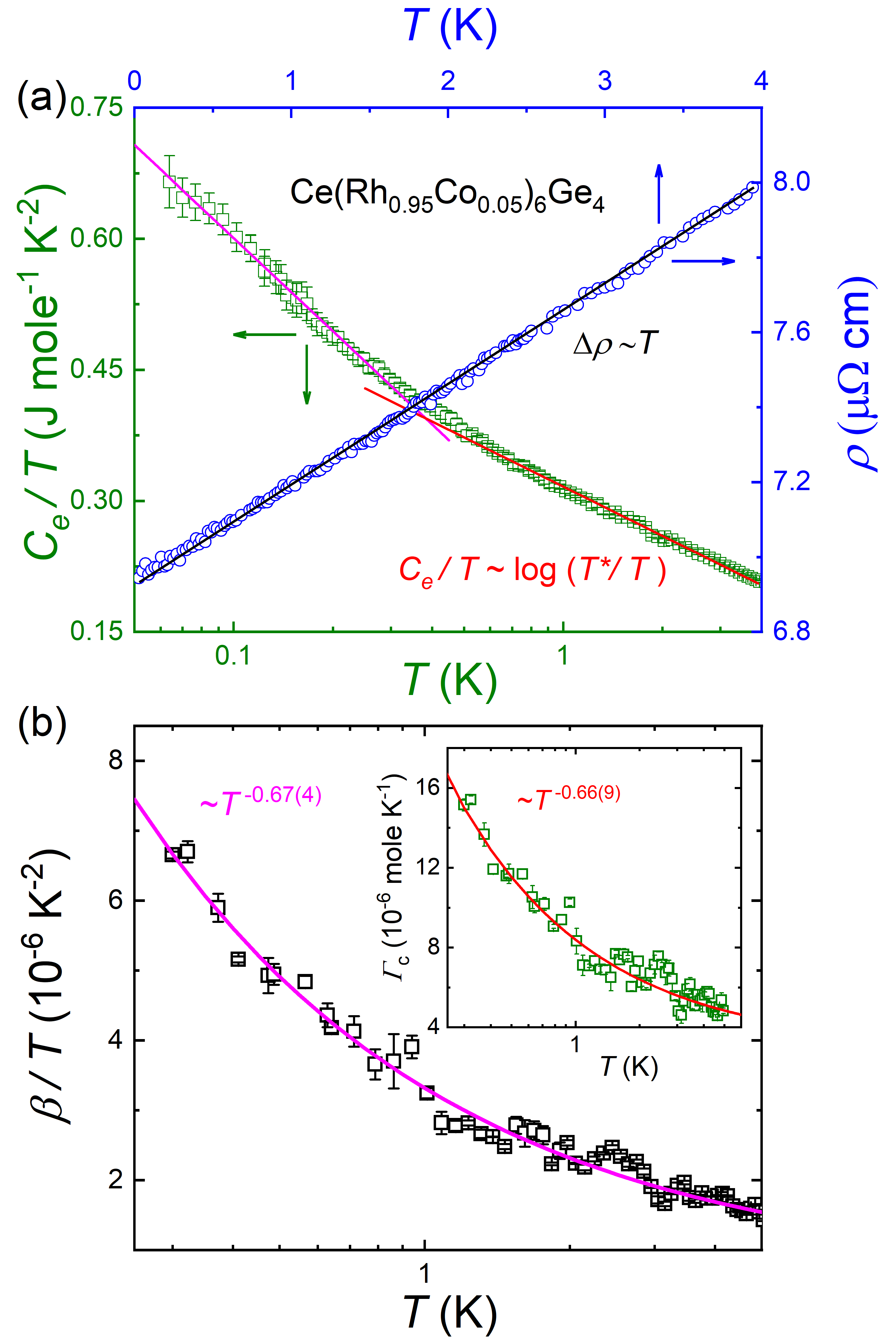}
	\end{center}
	\caption{(a) Temperature dependence of the specific heat coefficient (left axis) and resistivity (right axis) of a sample close to the critical concentration, Ce(Rh$_{0.95}$Co$_{0.05}$)$_6$Ge$_4$, measured down to 0.06~K. Clear strange metal behavior is observed, with the resistivity showing a linear-temperature dependence, while $C_e/T$ continues to increase with decreasing temperature,  following a logarithmic temperature dependence (red line), which exhibits a slope change below around 0.5~K. (b) Temperature dependence of the volume thermal expansion (as $\beta/T$), together with fit to a power law $\beta/T\sim T^{-n}$ shown by the solid line with $n=0.67(4)$. The inset shows the temperature dependence of the Grüneisen ratio $\beta/C_e$ that diverges at low temperatures, with an exponent $n=0.66(9)$.  }   \label{Fig4}\end{figure}

The behavior at the critical concentration $x_c\approx 0.05$ is highlighted in Fig.~\ref{Fig4}. In panel (a), $\rho(T)$ and the electronic specific heat coefficient $C_e/T$ are displayed down to 0.06~K. Here $\rho(T)$ remains linear down to the base temperature, mirroring the observation in pressurized \serge over almost two decades of temperature \cite{Shen2020}. Meanwhile $C_e/T$ exhibits a logarithmic divergence $C_e/T\sim{\rm log}(T^*/T)$  below around 4~K, where the data down to 0.5~K are described by a characteristic temperature \cite{Stewart2001} $T^*=48.9$~K (red solid line). At lower temperatures there is a change of slope leading to a more divergent $C_e/T$, which can be accounted for by either a logarithmic divergence with $T^*=5.0$~K, or a power law divergence $\sim T^{-0.31}$. A similar change of slope is also observed in CeRh$_6$Ge$_4$ under hydrostatic pressure (below which  $T^*=2.3~$~K is deduced) \cite{Shen2020}, in YbNi$_4$(P$_{0.92}$As$_{0.08}$)$_2$ \cite{Steppke2013} and in stoichiometric as well as 5\% Ge-substituted YbRh$_2$Si$_2$ \cite{YRSnature,YRSGruneisen}. At very low temperatures below 0.08~K, $C_e/T$ may slightly flatten. This could signal the emergence of a heavy Fermi liquid state (corresponding to a greatly enhanced Sommerfeld coefficient $\gamma\approx700$~mJ/mol~K$^2$), situating the sample very slightly above the QCP. 

Figure~\ref{Fig4}(b) shows the temperature dependence of   $\beta/T$, with the corresponding Gr\"uneisen ratio $\Gamma_{c}(T)=\beta(T)/C_e(T)$ shown in the inset. Clear divergences are observed in both quantities, demonstrating a QCP in \sergeCo \cite{Zhu2003,Kuechler2003,Gegenwart2016}, with best fits  $\beta(T)/T\sim T^{-0.67(4)}$, and $\Gamma_{c}(T)\sim T^{-0.66(9)}$. 

%discussion
At first sight, our findings of $\nu z=3/2$ inferred from the exponent  in $\Gamma_{c}(T)$ of  $\approx-2/3$  might be interpreted as the
observation of a classic Hertzian ferromagnetic QCP, with $z=3$  \cite{Zhu2003} (and $\nu=1/2$,  following \cite{Hertz1976}).  However, this interpretation cannot be reconciled with the strange metal behavior \cite{Shen2020}, for on the one hand a
Hertzian QCP gives rise to a $T^{5/3}$ resistivity \cite{Stewart2001} while on the other, it is inconsistent with the local moment character of the phase transition, evidenced by the large $S_0$ in the heat capacity $C/T=S_0/T^*\log(T^*/T)$ \cite{Shen2020}.  

 Moreover, CeRh$_6$Ge$_4$ is an easy-plane ferromagnet, in which the  critical magnetic modes lie in the basal plane, and on rather general grounds we do not expect $z=3$, because the in-plane magnetization is not conserved. We recall that in an itinerant ferromagnet, collective magnetic modes are overdamped with a dynamical susceptibility that has the form $\chi^{-1}(q,\omega)\sim m+v^2q^2+i\omega/D(q)+O(\omega^2)$, where $m$ is the quantum critical tuning parameter and $D(q)$ is the momentum-dependent decay rate of the magnetization at wave vector $q$ in the underlying electron sea. 
Away from criticality, $m>0$ is finite giving rise to a dynamical susceptibility with a Lorentzian lineshape with a width $\tau^{-1}\sim mD(q)$. If the magnetization is conserved, then the decay rate $D(q)\propto q$ vanishes linearly with momentum. However, in an easy-plane ferromagnet the magnetization is not conserved, so that 
$\tau^{-1}(q=0)\sim mD(0)$ is finite, leading to a constant rate of decay of the magnetization at $q=0$, away from criticality. At the QCP where $m$ vanishes, the decay rate $\tau^{-1}(q)\sim q^z\propto q^2D(q)$ now determines the dynamical critical exponent. Thus, if the magnetization is conserved $z=3$ \cite{Hertz1976,Sachdev2011}, but since an easy-plane magnetization is not conserved $\tau^{-1}(q)\sim q^2D(0)$ implies $z=2$.

Further symmetry considerations reinforce the departure in CeRh$_6$Ge$_4$ from the $z=3$ expected in a Hertzian model of ferromagnetism.  Here,  the  non-centrosymmetric crystal removes the Kramers degeneracy of the Fermi surface,  which as shown in the Supplemental material  \cite{SI} leads to an excitation gap for in-plane spin fluctuations of conduction electrons. This gives rise to an effective vacuum for the critical magnetic modes 
which become undamped, giving rise to a reduction of the dynamical critical exponent. For a complete gap, this leads to $z=1$, predicting $\Gamma_{c}\sim T^{-2}$, but in the presence of touching points between spin-split Fermi surfaces (the example treated in \cite{SI}), a residual damping leads to $z=2$, predicting $\Gamma_c\sim T^{-1}$. These are both in disagreement with the observed exponent. Together, these arguments rule out the magnetism as the origin of the underlying criticality in our experiment and in so doing implies  the criticality lies beyond the Landau paradigm. 

An appealing alternative scenario lies in a possible connection between the observed $z=3$ criticality and the  small-to-large Fermi surface transition of a strange metal \cite{SI}.  The quasiparticle charge density, described by the Fermi surface volume, undergoes a 
transition in a Kondo-breakdown QCP \cite{Schrder2000,Si2001,Coleman2001,Komijani2019}. Notably, while quantum oscillation experiments indicate localized $4f$ electrons at ambient pressure \cite{Wang2021}, an abrupt reconfiguration of the electronic structure in the vicinity of the QCP has been recently signalled in \cerge from thermopower measurements \cite{Thomas2024}.
The essence of the small to large Fermi surface transition
is an Anderson-Higgs process in which  the phase fluctuations of the hybridization between conduction and $f$-electrons are absorbed into the f-electron gauge field, giving them charge  \cite{Senthil03,Coleman05,Wugalter20}. Unlike a relativistic field theory where critical modes propagate at the speed of light, here they move within a dissipative Fermi sea, separating  into two components: amplitude modes, which like zero sound in a Landau Fermi liquid, lie above the particle-hole continuum, forming a $z=1$ mode, and gauge modes associated with local conservation of $f$-charge, which lie inside the particle-hole continuum, coupling to the conserved charge of the quasiparticles and giving rise to a $z=3$ criticality \cite{Pepin2007,Pepin2008}. The key point here is that phase fluctuations of the hybridization are over-damped through their coupling to charge via particle-hole excitations. Since the total charge density 
is a constant of motion, the over-damped dissipation rate vanishes linearly as $q\to 0$.
This leads to a dynamical susceptibility for hybridization fluctuations of the form  $\chi_\varphi^{-1}(q,\omega)\sim q^2+i\omega/q$. One can see from this that $\chi_\varphi\sim q^{-2}f(\omega/q^3)$, corresponding to $z=3$ dynamical scaling. Thus by ruling out magnetism, we are led to identify these soft phase fluctuations, mixed with gauge modes, as the most likely origin of the singular thermodynamics and the observed $z=3$ divergence of the Gr\"uneisen parameter.

This framework allows us to unify various recent observations about strange metal criticality. 
For instance, we are able to understand the observation of an almost identical divergence in the Gr\"uneisen parameter $\Gamma_{c}\sim T^{-0.7}$ in antiferromagnetic YbRh$_2$Si$_2$ \cite{Kuechler2003}, despite a completely different magnetic structure, as a natural consequence of the conserved dynamics of the small-to-large Fermi surface transition.
Based on our arguments, we expect a $z=3$ gauge mode to develop as a universal precursor of partial Mott localization, independently of the mechanism or whether the transition involves staggered antiferromagnetism \cite{Phillips2022,Husain2019}, 
ferromagnetism, valley \cite{Lu2019} or nematic \cite{Licciardello2019} order. More generally, these results underscore the pivotal role of charge-conserving modes \cite{Kimura2006,Prochaska2020,Kobayashi2023} in a diverse range of strange metal systems. A future measurement of the  Gr\" uneisen parameter at the 
strange metal nematic critical point of sulphur-doped
FeSe$_{1-x}$S$_{x}$ \cite{Licciardello2019} will provide an important test of these ideas.

\begin{acknowledgements}
This work was supported by the National Key R\&D Program of China (Grant No. 2022YFA1402200 and No. 2023YFA1406303),  the National Natural Science Foundation of China (Grants No. 12034017, No. W2511006, No. 12222410, No. 12174332, No. U23A20580, No. 12350710785, and No. 12204159) and the  National Science Foundation of the United States of America, grant DMR-1830707 (Piers Coleman). 

\end{acknowledgements}

\end{document}